\documentclass[12pt,a4paper]{article}
\usepackage[latin1]{inputenc}
\usepackage[T1]{fontenc}
\usepackage[english]{babel}
\usepackage{lmodern}
\usepackage{hyperref}
\usepackage[dvips]{graphics,color}
\usepackage{relsize}
\usepackage{cancel}
\usepackage{amsmath,amssymb,amsfonts}
\usepackage{bm}
\usepackage{tikz}
\title{Quantum dissipative effects for a real scalar field  coupled to 
a dynamical Neumann surface in $d+1$ dimensions}
\author{C.~D.~Fosco \\
and \\
B.~C.~ Guntsche\\
{\normalsize\it Centro At\'omico Bariloche and Instituto Balseiro}\\
{\normalsize\it Comisi\'on Nacional de Energ\'\i a At\'omica}\\
{\normalsize\it R8402AGP Bariloche, Argentina.}}
\begin{document}
\date{}
\maketitle
\begin{abstract}
We study dissipative effects for a system consisting of a massless real 
scalar field satisfying Neumann boundary conditions on a space and 
time-dependent surface, in $d+1$ dimensions. We focus on the comparison of 
the results for this system with the ones corresponding to Dirichlet 
conditions, and the  same surface space-time geometry. 
We show that, in $d=1$, the effects are equal up to second order for 
rather arbitrary surfaces, and up to fourth order for wavelike surfaces. 
For $d>1$, we find general expressions for their difference. 
\end{abstract}
\section{Introduction}\label{sec:intro}
Many interesting macroscopic effects are a result of the interaction of an 
object which can be accurately described classically, with 
a quantum field, like the Casimir 
effect~\cite{milton1999casimir,bordag2009advances}.
 Among those effects, not the least important is the so 
called `motion induced radiation', or Dynamical Casimir Effect (DCE)
\cite{Moore,FulDav,reviews,Good,Cong,Fulling}, 
whereby radiation is emitted when an object imposing boundary conditions 
on a quantum field, is accelerated.  The quantum nature of the phenomenon 
is manifested, for instance, in that the expectation value of the
radiation field, linear in the sources, vanishes, while observables 
(quadratic in the radiation field) do not.  Indeed, it is the
{\em correlation\/} between the fluctuations in the sources, mediated by 
the quantum field, what makes the effect possible.

In the study of dissipative phenomena, a useful method involves examining 
the imaginary part of the in-out effective action, which is a functional 
of the external fields. For instance, in the case of motion-induced 
radiation, these fields are variables that represent the time-varying 
geometry of the system in which the fields are present.
A path integral approach has been successfully used to study this kind of 
system~\cite{Golestanian:1997ks,Golestanian:1998bx} and a closely related 
subject, namely, radiation from time-dependent boundary 
conditions with a fixed geometry~\cite{Karabali:2015epa}.

Once obtaining an expression for the effective action, dissipation 
manifests itself in the existence of non-analyticities, when the external 
fields are Fourier transformed to momentum space.
A consequence of this is that one may discard analytic terms when 
evaluating the dissipative effects, a procedure which greatly simplifies 
the calculations.  

In a previous paper~\cite{Fosco:2024bqx}, we carried out this study for 
the dynamical Casimir effect due to a real scalar field in $d+1$
dimensions, in the presence of a mirror that imposes Dirichlet boundary 
conditions and undergoes time-dependent motion or deformation. Using a 
perturbative approach, we expanded in powers of the deviation of the 
mirror's surface from a hyperplane, up to fourth order. Here we 
find expressions for the difference between the Neumann and Dirichlet 
cases. We explicitly show some interesting phenomena, like the vanishing 
of the difference between Neumann and Dirichlet for $d=1$, and the fact 
that a Neumann condition is, for the same geometry, more effective in 
producing pairs than a  Dirichlet one.

The structure of this paper is as follows: in Sect.~\ref{sec:thesystem} we 
define the system, and introduce the object we plan to evaluate, comparing 
it to its Dirichlet boundary conditions analogue. Then, in 
Sect.~\ref{sec:pert} we evaluate the effective action, focusing on its 
imaginary part, within the context of a perturbative expansion in powers 
of the departure of the Neumann surface with respect to a planar one. We 
do that up to the fourth order in that expansion.
Finally, in Sect.~\ref{sec:conc}, we present our conclusions.

\section{The system and its effective action}\label{sec:thesystem}
We study a system consisting of a massless real scalar field $\varphi(x)$
in \mbox{$d+1$} dimensions, subjected to  Neumann boundary conditions on a
space-time surface $\Sigma$, but otherwise described by the free action
${\mathcal S}_0(\varphi)$:
\begin{equation}\label{eq:defs0}
{\mathcal S}_0(\varphi) \,=\, \frac{1}{2} \int d^{d+1}x \, 
\partial_\mu \varphi(x)  \partial_\mu \varphi(x) \;.
\end{equation}
We use letters from the middle of the Greek alphabet ($\mu, \nu, \ldots$)
to design space-time indices, running over the values $0, 1, \ldots, d$;
the $0$ index being reserved for the temporal components of an object.
Space-time coordinates are $x^\mu = x_\mu$, $x_0$ being the imaginary time
(we use natural units, such that: $\hbar \equiv 1$ and $c \equiv 1$). Since
the metric tensor becomes the $(d+1) \times (d+1)$ identity matrix, no
meaning is to be ascribed to the position (upper or lower) of a particular
index.  Finally,
Einstein convention of summation over repeated indices in monomial
expressions is also assumed, unless explicitly stated otherwise.  

A convenient way to describe the system, while taking into account the 
boundary conditions, is through the Euclidean vacuum transition
amplitude, a functional of $\Sigma$:
\begin{equation}\label{eq:defzn}
{\mathcal Z}(\Sigma) \;=\; \int {\mathcal D}\varphi \,
\delta_\Sigma(\partial_n\varphi) \, e^{- {\mathcal S}_0(\varphi)} \;,
\end{equation}
where $\delta_\Sigma(\partial_n \varphi)$ denotes a Dirac $\delta$ 
functional, which imposes Neumann conditions for $\varphi$ on the surface. 

We shall now be more specific about the kind of surface we deal with here, 
since our treatment relies upon some assumptions regarding
that object. Indeed, we assume it to be possible, at least by a proper 
coordinate system choice, to parametrize $\Sigma$ by $y^\mu$ $(\mu = 0, 
\ldots , d)$ as follows:
\begin{align}\label{eq:defmonge}
\Sigma )\;\;  & x_\shortparallel \;\to\; y(x_\shortparallel)
\;\;, \;\;\; x_\shortparallel \equiv (x^\alpha)_{\alpha=0}^{d-1}
\;\;\; {\rm and:}
\nonumber\\
& y^\mu(x_\shortparallel) \,=\, \delta^\mu_\alpha x^\alpha \,+\, 
\delta^\mu_d \, \psi(x_\shortparallel) \;\;. 
\end{align}
This allows us to represent $\delta_\Sigma(\partial_n \varphi)$ more 
explicitly, by introducing an auxiliary field $\lambda(x_\shortparallel)$:
\begin{equation}\label{eq:aux1}
\delta_\Sigma(\partial \varphi) \;=\; \int {\mathcal D}\lambda \, e^{i \int
d^dx_\shortparallel \sqrt{g(x_\shortparallel)} \, 
\lambda(x_\shortparallel)\,{n}_\mu(x_\shortparallel) \partial_\mu 
\varphi(x_\shortparallel, \psi(x_\shortparallel)) } \;,
\end{equation}
where: $g(x_\shortparallel) \equiv {\rm
det}[g_{\alpha\beta}]$, with  $g_{\alpha\beta} = \delta_{\alpha\beta}+
\partial_\alpha\psi \partial_\beta\psi$, the induced metric tensor on
$\Sigma$, and $n_\mu(x_\shortparallel)$ the unit normal vector:
\begin{equation}
n^\mu(x_\shortparallel) \;=\; \frac{N^\mu
(x_\shortparallel)}{||N(x_\shortparallel)||} \;,\;\; N^\mu
(x_\shortparallel) \,=\, \delta^\mu_d - 
\delta^\mu_\alpha \partial_\alpha \psi(x_\shortparallel) \;.
\end{equation}
The factor $\sqrt{g(x_\shortparallel)}$ in (\ref{eq:aux1}),
introduced in order to have a reparametrization invariant expression, is 
canceled by $||N(x_\shortparallel)||$, with which it coincides. 

The effective action $\Gamma(\Sigma)$, which, given the specific
parametrization (\ref{eq:defmonge}), we also denote by $\Gamma(\psi)$, is
obtained by functionally integrating out the quantum field:
\begin{equation}\label{eq:defgamma_1}
e^{-\Gamma(\Sigma)} \;=\; e^{-\Gamma(\psi)} \;=\; \frac{{\mathcal 
Z}(\Sigma)}{{\mathcal Z}_0} 
\;\;, \;\; {\mathcal Z}_0 \,\equiv\, \int {\mathcal D}\varphi \, e^{-
{\mathcal S}_0(\varphi)} \;.	
\end{equation}
Integrating out $\varphi$, we obtain for $\Gamma$ an
expression~\footnote{We use the shorthand notation: $\int d^{d+1}x \ldots
\equiv \int_x \ldots \;\;, \int d^dx_\shortparallel \ldots  \equiv
\int_{x_\shortparallel} \ldots \;\;, \;\; \ldots$}  as a functional
integral over $\lambda$:
\begin{equation}\label{eq:eff_4}
e^{-\Gamma(\psi)} \,=\, \int \mathcal{D}\lambda \; 
e^{-\frac{1}{2} \int_{x_\shortparallel, x'_\shortparallel} 
\sqrt{g(x_\shortparallel)} \lambda(x_\shortparallel) \,  
\sqrt{g(x'_\shortparallel)} \lambda(x'_\shortparallel) \,
\langle \chi(x) \, \chi(x') \rangle \big|_{\substack{x_d =
\psi(x_\shortparallel) \\ x'_d = \psi(x'_\shortparallel)}}}
\end{equation}
where a new field $\chi$ has been defined (for all $x$) such that
\mbox{$\chi(x) \equiv n_\mu(x_\shortparallel) \partial_\mu \varphi(x)$}, 
and: 
\begin{equation}
\langle \ldots \rangle \;\equiv\; \frac{\int {\mathcal D}\varphi
\ldots e^{- {\mathcal S}_0(\varphi)}}{{\mathcal Z}_0} \;.
\end{equation}

From the effective action in its (\ref{eq:eff_4}) representation, it
becomes evident its similarity to the one for $\Gamma_D(\psi)$ the 
effective action for {\em Dirichlet\/} boundary conditions:
\begin{equation}\label{eq:eff_d}
e^{-\Gamma_D(\psi)} \,=\, \int \mathcal{D}\lambda \; 
e^{-\frac{1}{2} \int_{x_\shortparallel, x'_\shortparallel} 
\sqrt{g(x_\shortparallel)} \lambda(x_\shortparallel) \,  
\sqrt{g(x'_\shortparallel)} \lambda(x'_\shortparallel) \,
\langle \varphi(x) \, \varphi(x') \rangle \big|_{\substack{x_d =
\psi(x_\shortparallel) \\ x'_d = \psi(x'_\shortparallel)}}} \;.
\end{equation}
Indeed, (\ref{eq:eff_4}) may be interpreted as the effective action, with 
Dirichlet boundary conditions on $\Sigma$, but with the replacement:  
$\varphi(x) \to \chi(x)$. This property suggests the possibility of 
comparing the effective actions for Neumann and Dirichlet boundary 
conditions, to find out the nature and properties of the difference.

To proceed with the calculation of $\Gamma$, we now integrate out 
$\lambda$, finding as a result:
\begin{align}\label{eq:eff_5}
\Gamma(\psi) &=\; \frac{1}{2} \, 
{\rm log} \det\big[ K(x_{\shortparallel},x'_{\shortparallel})\big]
\nonumber\\ 
K(x_{\shortparallel},x'_{\shortparallel}) & \equiv\, 
N_\mu(x_\shortparallel) \,
\big[\Delta_{\mu\nu}(x - x')\big]\Big|_{\substack{x_d =
\psi(x_\shortparallel) \\ x'_d = \psi(x'_\shortparallel)}}
\, N_\nu(x'_\shortparallel) 
\end{align}
with
\begin{equation} \label{def: propagator}
\Delta_{\mu\nu}(x - x') \,\equiv\, \partial_\mu \partial'_\nu \Delta(x-x')
\;,\;\;\; \Delta(x-x') \,\equiv\,\int_{\not{k}}
\frac{e^{ik(x-x')}}{k^2} \;,
\end{equation}
($\int_{\not{k}} \equiv (2 \pi)^{-(d+1)}\int_k$).

Thus, what remains is to evaluate the log of the functional determinant of 
the kernel $K$ or, equivalently:
\begin{equation} \label{eq:eff_6}
\Gamma(\psi) \;=\;
\frac{1}{2} {\rm Tr}\big\{\log\big[K(x_{\shortparallel},x'_{\shortparallel})
\big]\Big\} \;.
\end{equation}
From its very definition, we obtain for the kernel an expression which may 
be conveniently rendered as follows:
\begin{equation}\label{eq:KNeu}
K\;=\;K_S + K_T + K_U \;,
\end{equation}
with
\begin{align}\label{eq:defks}
K_S(x_\shortparallel,x'_\shortparallel) \; = & 
\Delta_{dd}\big(x_\shortparallel-x'_\shortparallel,\psi(x_\shortparallel)
 -  
\psi(x'_\shortparallel) \big)
\nonumber\\
K_T(x_\shortparallel,x'_\shortparallel) \; = & - 
\Delta_{d\alpha}\big(x_\shortparallel-x'_\shortparallel,\psi(x_\shortparallel)
-  \psi(x'_\shortparallel)\big) \partial_\alpha\psi(x'_\shortparallel) 
\nonumber\\
- & \partial_\alpha \psi(x_\shortparallel) \Delta_{\alpha 
d}\big(x_\shortparallel-x'_\shortparallel,\psi(x_\shortparallel) -  
\psi(x'_\shortparallel)\big) \nonumber\\
 K_U(x_\shortparallel,x'_\shortparallel) \; = & 
 \partial_\alpha \psi(x_\shortparallel) \Delta_{\alpha 
 \beta}\big(x_\shortparallel-x'_\shortparallel,\psi(x_\shortparallel) -  
 \psi(x'_\shortparallel)\big) \partial_\beta\psi(x'_\shortparallel) \,.
\end{align}

\section{Expansion in powers of $\psi$}\label{sec:pert}
We evaluate $\Gamma$ by expanding it in powers of $\psi$. To that end, we 
first expand $K(x_\shortparallel,x'_\shortparallel)$,
\begin{equation}
	K(x_\shortparallel,x'_\shortparallel) \;=\; 
	K^{(0)}(x_\shortparallel,x'_\shortparallel) \,+\, 
	K^{(2)}(x_\shortparallel,x'_\shortparallel) \,+\, \ldots 
\end{equation}
where the index denotes order in $\psi$, and we have incorporated the fact 
that there are only even powers in the expansion of the kernel.

\subsection{Second order}
When expanding in powers of $\psi$, the departure of $\Sigma$ with respect 
to a plane, we shall treat the contributions coming from the three terms 
in (\ref{eq:defks}) above in turn. In order to calculate the second 
order term, we see that
\begin{equation}
	\Gamma^{(2)}(\psi) \;=\; \frac{1}{2} {\rm Tr} \big\{ [K^{(0)}]^{-1}\, 
	K^{(2)} \big\}
\end{equation}
so that, taking into account (\ref{eq:defks}), there will be three 
contributions in $\Gamma^{(2)}(\psi)$
\begin{equation}
\Gamma^{(2)}(\psi) \;=\; \Gamma^{(2)}_S(\psi) \,+\,\Gamma^{(2)}_T(\psi) 
\,+\,\Gamma^{(2)}_U(\psi) \;,
\end{equation}
in an obvious notation.

It is convenient in what follows to interpret the kernels above as matrix 
elements of operators. In a Dirac bra-ket notation, we have:
\begin{equation} \label{deff: operators}
K^{(0)}(x_\shortparallel,x'_\shortparallel) \,=\, \langle x_\shortparallel | 
\widehat{K}^{(0)} | x'_\shortparallel \rangle 
 \,=\, - \, \int_{\not{p_\shortparallel}} 
 e^{ip_\shortparallel \cdot (x_\shortparallel-x'_\shortparallel)}
\frac{|\widehat{p}_\shortparallel|}{2} \;.
\end{equation}
Hence,
\begin{equation} 
	\widehat{K}^{(0)} \,=\,	- \, \frac{|\widehat{p}_\shortparallel|}{2}
\end{equation}
with $|\widehat{p}_\shortparallel| = \sqrt{\widehat{p}_\alpha 
\widehat{p}_\alpha}$, 
and $\langle x_\shortparallel|\widehat{p}_\alpha |x'_\shortparallel \rangle 
= -i \partial_\alpha \delta(x_\shortparallel - x'_\shortparallel)$.
To functions of $x_\shortparallel$ there correspond operators which are
diagonal in this representation. For example, to the departure $\psi$ 
there corresponds an operator $\widehat{\psi}$, with \mbox{$\langle 
x_\shortparallel|\widehat{\psi}|x'_\shortparallel \rangle 
= \psi(x_\shortparallel) \delta(x_\shortparallel - x'_\shortparallel)$}.
Besides, we have the fundamental commutation relation
$[ \widehat{x}_\alpha , \widehat{p}_\beta ] = i \delta_{\alpha\beta}$.

Now take, for instance, the operator $K_S$, for which we know  
$K^{(2)}_S(x_\shortparallel, x'_\shortparallel)$ by expanding 
\eqref{eq:eff_5}:
\begin{equation}\label{deff K^2 1}
 K^{(2)}_S(x_\shortparallel, x'_\shortparallel) = -\int_{\not 
 p_\shortparallel} 
 e^{ip_\shortparallel(x_\shortparallel-x'_\shortparallel)} 
 \frac{|p_\shortparallel|^3}{4} (\psi(x_\shortparallel)^2 
 +\psi(x'_\shortparallel)^2 -2\psi(x_\shortparallel) 
 \psi(x'_\shortparallel))\,.
\end{equation}
We want to build the operator $\widehat{K_S}^{(2)}$ so that the 
expression above corresponds to $\langle x_\shortparallel | 
\widehat{K_S}^{(2)}|x'_\shortparallel\rangle$. To do so, we insert 
the identity in the $x_\shortparallel$ and $p_\shortparallel$ bases, i.e 
$\int_{x_\shortparallel}| x_\shortparallel\rangle \langle 
x'_\shortparallel|$ and $\int_{p_\shortparallel} | p_\shortparallel\rangle 
\langle p'_\shortparallel|$, use $\widehat{\psi}\,|x_\shortparallel\rangle 
= \psi(x_\shortparallel)\,|x_\shortparallel\rangle$ and 
$|\widehat{p_\shortparallel}|\,|p_\shortparallel\rangle = 
|p_\shortparallel|\,|p_\shortparallel\rangle$, and exploit the relation 
$\langle x_\shortparallel|p_\shortparallel\rangle = 
e^{ip_\shortparallel \cdot x_\shortparallel}/(2\pi)^{d/2}$. 

Then we see that:
\begin{equation} 
	\widehat{K}_S^{(2)} \,=\,	-\frac{1}{4} \,\big( \widehat{\psi}^2  \, 
	|\widehat{p}_\shortparallel|^3\,+\, |\widehat{p}_\shortparallel|^3 
 \,\widehat{\psi}^2 \big)\,+\,\frac{1}{2} \, \widehat{\psi}  \, 
	|\widehat{p}_\shortparallel|^3 \, \widehat{\psi}  \;,
\end{equation}
therefore,
\begin{align}
	\Gamma^{(2)}_S(\psi) &=\, \frac{1}{4} \,{\rm Tr}\big(
	|\widehat{p}_\shortparallel|^{-1} \, \widehat{\psi}^2 
	|\widehat{p}_\shortparallel|^3 \big) 
	\,+\, \frac{1}{4} \,{\rm Tr}\big(
	|\widehat{p}_\shortparallel|^{-1} \,	|\widehat{p}_\shortparallel|^3 
	\widehat{\psi}^2 \big) \nonumber\\
	 &-\frac{1}{2} \,{\rm Tr}\big(
	|\widehat{p}_\shortparallel|^{-1} \, \widehat{\psi} 
	|\widehat{p}_\shortparallel|^3 \, \widehat{\psi}
	\big) \;.
\end{align}
The first two terms lead to the same result, a  divergent contribution, 
which does not affect the imaginary part of the 
effective action. Indeed,
\begin{align}
&\frac{1}{4} \,{\rm Tr}\big(
|\widehat{p}_\shortparallel|^{-1} \, \widehat{\psi}^2 
|\widehat{p}_\shortparallel|^3 \big) 
\,+\, \frac{1}{4} \,{\rm Tr}\big(
|\widehat{p}_\shortparallel|^{-1} \,	|\widehat{p}_\shortparallel|^3 
\widehat{\psi}^2 \big) 
\nonumber\\
&=  \frac{1}{2} \,\int_{\not{k_\shortparallel}} |k_\shortparallel|^2 \,
\int_{x_\shortparallel} \, \psi^2(x_\shortparallel)  
\end{align}
which corresponds to an infinite renormalization of the mass of the 
surface. Discarding this term, we are lead to: 
\begin{equation}
\Gamma^{(2)}_S(\psi) \,=\, -\frac{1}{2} \,{\rm Tr}\big(
|\widehat{p}_\shortparallel|^{-1} \, \widehat{\psi} 
|\widehat{p}_\shortparallel|^3 \, \widehat{\psi}
\big) \;.
\end{equation}
We now proceed to write the term above in an equivalent way, by noticing 
that:
\begin{align}
\Gamma^{(2)}_S(\psi) &=\, -\frac{1}{2} \,{\rm Tr}\big(
|\widehat{p}_\shortparallel|^{-1} \, \widehat{\psi} \,
\widehat{p}_\shortparallel^2\, |\widehat{p}_\shortparallel| \, 
\widehat{\psi}
\big) \nonumber\\
&=\, -\frac{1}{2} \,{\rm Tr}\big(
|\widehat{p}_\shortparallel| \, \widehat{\psi} \,
|\widehat{p}_\shortparallel| \, 
\widehat{\psi}
\big) \,-\,\frac{1}{2} \,{\rm Tr}\big(
|\widehat{p}_\shortparallel|^{-1} \,[ \widehat{\psi}\,,\,
\widehat{p}_\shortparallel^2] \,
|\widehat{p}_\shortparallel| \, 
\widehat{\psi}\big) \;.
\end{align}

We recall now the Dirichlet case, where one can write the (full) second 
order contribution to the effective action, 
$\Gamma_D^{(2)}$, as follows:
\begin{equation}
	\Gamma^{(2)}_D(\psi) \,=\, -\frac{1}{2} \,{\rm Tr}\big(
	|\widehat{p}_\shortparallel| \, \widehat{\psi} 
	|\widehat{p}_\shortparallel| \, \widehat{\psi}
	\big) \;,
\end{equation}
i.e.,
\begin{equation}
	\Gamma^{(2)}_S(\psi) \,=\, \Gamma^{(2)}_D(\psi) \,-\,\frac{1}{2} 
	\,{\rm Tr}\big(
	|\widehat{p}_\shortparallel|^{-1} \,[ \widehat{\psi}\,,\,
	\widehat{p}_\shortparallel^2] \,
	|\widehat{p}_\shortparallel| \, 
	\widehat{\psi}\big) \;.
\end{equation}
After some algebra, we may render the above expression as follows:
\begin{align}
\Gamma^{(2)}_S(\psi) \,-\, \Gamma^{(2)}_D(\psi) \,& =\, -\,
{\rm Tr}\big(	|\widehat{p}_\shortparallel|^{-1} \,\hat{\partial}_\alpha 
\,  (\partial_\alpha \widehat{\psi}) \,|\widehat{p}_\shortparallel| \, 
\widehat{\psi}\big) \nonumber\\
& - \, \frac{1}{2} \,{\rm Tr}\big(	|\widehat{p}_\shortparallel|^{-1}  \,  
(\partial_\alpha \widehat{\psi}) 
\,|\widehat{p}_\shortparallel| \, (\partial_\alpha \widehat{\psi})
\big) \;.
\end{align}
A word about notation: when we put parenthesis on a derivative of the 
operator $\widehat{\psi}$, that should be understood as the operator 
corresponding to the derivative of the function. Otherwise, a 
derivative means an operator, and is explicitly denoted as such.

For the other two kernels, we have:
\begin{equation} 
	\widehat{K}_T^{(2)} \,=\, \frac{i}{2} \big((\partial_\alpha\widehat{\psi}) 
	\widehat{\psi}  
	\,	|\widehat{p}_\shortparallel| \widehat{p}_\alpha 
	\,-\,   (\partial_\alpha\widehat{\psi}) 
	|\widehat{p}_\shortparallel| 
	\widehat{p}_\alpha   \widehat{\psi}+  
	\widehat{\psi}  
	\,	|\widehat{p}_\shortparallel| \widehat{p}_\alpha (\partial_\alpha\widehat{\psi})
	\,-\,  
	|\widehat{p}_\shortparallel| 
	\widehat{p}_\alpha   \widehat{\psi}  (\partial_\alpha\widehat{\psi})  \big)	 \;,
\end{equation}
and
\begin{equation} 
	\widehat{K}_U^{(2)} \,=\, \frac{1}{2} 
	(\partial_\alpha\widehat{\psi}) 
	\frac{\widehat{p}_\alpha\widehat{p}_\beta}{|\widehat{p}_\shortparallel|}
	(\partial_\beta \widehat{\psi}) \;,
\end{equation}
leading to  (discarding  terms which do not contribute to the imaginary 
part)
\begin{equation}
	\Gamma^{(2)}_T(\psi) \,=\,{\rm Tr}\big[
	|\widehat{p}_\shortparallel|^{-1} \, 
	(\partial_\alpha\widehat{\psi}) 
	|\widehat{p}_\shortparallel| \, (\partial_\alpha \widehat{\psi})	
	\big] \,+\,{\rm Tr}\big[
|\widehat{p}_\shortparallel|^{-1} \, \hat{\partial}_\alpha\, 
(\partial_\alpha\widehat{\psi}) 
|\widehat{p}_\shortparallel| \, \widehat{\psi}	
\big] \;,
\end{equation}
and 
\begin{equation}
	\Gamma^{(2)}_U(\psi) \,=\,- \frac{1}{2} \, {\rm Tr}\big[
	|\widehat{p}_\shortparallel|^{-1} \, 
	(\partial_\alpha\widehat{\psi}) \,  \frac{\widehat{p}_\alpha 
		\widehat{p}_\beta}{|\widehat{p}_\shortparallel|}  \, 
	(\partial_\beta\widehat{\psi})\Big]
	\;.
\end{equation}

Summing up the three contributions, we see that the result for 
$\Gamma^{(2)}$ may be put as follows:
\begin{align}
\delta\Gamma^{(2)}(\psi) & \equiv \Gamma^{(2)}(\psi) - 
\Gamma_D^{(2)}(\psi) \nonumber\\
& =	\frac{1}{2}  {\rm Tr}\big[
	|\widehat{p}_\shortparallel|^{-1} \, 
	(\partial_\alpha\widehat{\psi}) |\widehat{p}_\shortparallel|
\big( \delta_{\alpha\beta} - \frac{\widehat{p}_\alpha 
\widehat{p}_\beta}{	
|\widehat{p}_\shortparallel|^2} \big)  \, 
(\partial_\beta\widehat{\psi})	
	\big] \,.
\end{align}
By using the momentum and coordinate representations judiciously, the 
expression above may be converted into:
\begin{equation} \label{def: dif order 2}
	\delta\Gamma^{(2)}(\psi) \,=\,
	\frac{1}{2} \int_{\not{k_\shortparallel}} \, F(k_\shortparallel) \, 
	\big| \tilde{\psi}(k_\shortparallel) 
	\big|^2 \;,
\end{equation}
where
\begin{equation}\label{eq:fk}
F(k_\shortparallel) \,=\,	k_\shortparallel^2 \, 
\int_{\not{p_\shortparallel}}
	\frac{|p_\shortparallel|}{|p_\shortparallel+k_\shortparallel|}
	\Big[ 1 - \big(\frac{p_\alpha  k_\alpha}{|p_\shortparallel| 
	\,|k_\shortparallel| }\big)^2 \Big] \;.	
\end{equation}

When $d=1$, we note that the integration variable has just one component 
(a frequency). Therefore
the integrand vanishes, and the difference between Neumann and 
Dirichlet cases disappears at this order for $d=1$.

Let us consider now (\ref{eq:fk}), for an arbitrary $d > 1$. We see that
 
\begin{equation}\label{eq:fk1}
	F(k_\shortparallel) \,=\, 
	\int_{\not{p_\shortparallel}} 
	\frac{p_\shortparallel^2 k_\shortparallel^2 - (p_\shortparallel \cdot  
	k_\shortparallel)^2}{|p_\shortparallel| \, |p_\shortparallel + 
	k_\shortparallel|}
	\;.	
\end{equation}
The integral may be exactly evaluated for any $d$, leading to:
\begin{equation}\label{eq:fk2}
	F(k_\shortparallel) \,=\, 
(k_\shortparallel^2)^{\frac{d+2}{2}} \, (d-1) \,
\frac{\left( \Gamma\left( \frac{d+1}{2} \right) \right)^2 
	\Gamma(-\frac{d}{2})}{
	2^{d+1} \pi^{\frac{d+2}{2}} \Gamma(d+1) }\,
\end{equation}
which allows us to compare with the Dirichlet case directly, both for odd 
and even $d$, by introducing~\eqref{eq:fk2} into~\eqref{def: dif order 2}.
\subsubsection{Odd $d$ $(d=2q+1)$}
Analyzing the case $d=2q+1$ for $q>0$ (since we have already established 
$\delta \Gamma^{(2)}(\psi)=0$ for $d=1$), we notice that the arguments of 
the involved Gamma functions are all well defined without the need of 
dimensional regularization, and there is no pole term. The imaginary part 
of the effective action in Minkowski signature, $\Gamma^{(2)}_M$, comes 
from the factor $(k_\shortparallel^2)^{\frac{d+2}{2}}$, which is not 
analytic in $k_\shortparallel^2$, so after the Wick rotation involving 
$k_\shortparallel^2\to - k_\shortparallel^2$, 
$i\Gamma^{(2)}_M=-\Gamma^{(2)}$, and $\int_{k_\shortparallel} \to 
-i\int_{k_\shortparallel}$, this factor develops an imaginary part when 
$k_\shortparallel^2>0$, i.e, the surface is described by timelike modes. 

\begin{figure}[h!] 
     \centering
     \includegraphics[width=0.7\textwidth]{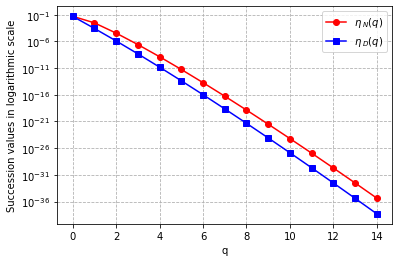}
     \caption{Successions $\eta_D$ and $\eta_N$ in logarithmic scale.}
     \label{fig:odd_ratio_order2}
\end{figure}

Writing $\text{Im}(\Gamma^{(2)}_M)$ for $q=\frac{d-1}{2}$ as

\begin{align} \label{def: imaginary neumann odd order 2}
    \text{Im}(\Gamma^{(2)}_M)= \frac{1}{2}(\eta_D+\eta_{\delta}) \int_{\not k_\shortparallel} \tilde{\psi}(k_\shortparallel) \tilde{\psi}(-k_\shortparallel) (k_\shortparallel^2)^{q+1} \sqrt{(k_0)^2-|\vec{k_\shortparallel}|^2} \;\Theta(|k_0|-|\vec{k_\shortparallel}|)\,
\end{align}
where $\eta_D(q)$ corresponds to the contribution in the Dirichlet case and $\eta_\delta(q)$ is the contribution of \eqref{eq:fk2}, we obtain the succession $\eta_N(q)=\eta_D(q)+\eta_\delta(q)$ for the Neumann case, which is compared to $\eta_D(q)$ in figure \ref{fig:odd_ratio_order2}.

\subsubsection{Even $d$ $(d=2q)$}

\begin{figure}[h!] 
     \centering
     \includegraphics[width=0.7\textwidth]{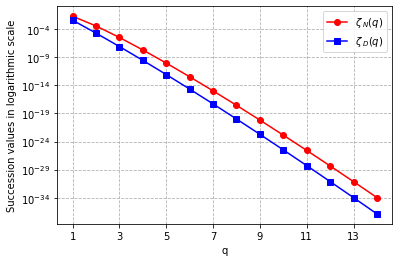}
     \caption{Successions $\zeta_D$ and $\zeta_N$ in logarithmic scale.}
     \label{fig:even_ratio_order2}
\end{figure}

For $d=2q$, the factor $\Gamma(-\frac{d}{2})$ yields an infinite counter-term analytic in $k_\shortparallel^2$ when using dimensional regularization $d=2q-\epsilon$ and taking the limit $\epsilon \to 0$. After the Wick rotation, the imaginary part of $\Gamma^{(2)}_M$ comes from the expansion $(k_\shortparallel^2)^{-\frac{\epsilon}{2}}=1-\frac{\epsilon}{2} \log(k_\shortparallel^2)$, which develops an imaginary part when $k_\shortparallel$ is timelike. This allows us to write

\begin{align} \label{def: imaginary neumann even order 2}
    \text{Im}(\Gamma^{(2)}_M)= \frac{1}{2}(\zeta_D+\zeta_{\delta}) \int_{\not k_\shortparallel} \tilde{\psi}(k_\shortparallel) \tilde{\psi}(-k_\shortparallel) (k_\shortparallel^2)^{q+1} \;\Theta(|k_0|-|\vec{k_\shortparallel}|)\,
\end{align}
where in a similar notation to the odd $d$ case, $\zeta_D$ is the contribution from the Dirichlet case, $\zeta_\delta$ is the contribution from \eqref{eq:fk2}, and we define $\zeta_N=\zeta_D+\zeta_\delta$, which is compared with $\zeta_D$ in figure \ref{fig:even_ratio_order2}\,.

 As we see for both odd and even dimensions, the Neumann boundary condition is more effective than the Dirichlet boundary condition when it comes to pair creation, and the probability decreases exponentially with growing dimensions in both cases.

 \subsection{Fourth order}

For this order, we have 

\begin{align}
    \Gamma^{(4)}(\psi) &=\;  \frac{1}{2} {\rm Tr}\big[[K^{(0)}]^{-1}K^{(4)}\big]   
\,-\,\frac{1}{4} {\rm Tr}\big[[K^{(0)}]^{-1}K^{(2)} [K^{(0)}]^{-1}K^{(2)}\big]\\ \nonumber
&= \Gamma^{(4,1)}(\psi) + \Gamma^{(4,2)}(\psi)\,
\end{align}
where we shall treat $\Gamma^{(4,1)}(\psi)$ and $\Gamma^{(4,2)}(\psi)$ separately.

\subsubsection{$\Gamma^{(4,1)}$}

For this contribution, we have

\begin{align}
    \widehat{K_S}^{(4)}&=-\frac{1}{48}\big(\widehat{\psi}^4 |\widehat{p_\shortparallel}|^5 +|\widehat{p_\shortparallel}|^5  \widehat{\psi}^4 \big) + \frac{1}{12} \big(\widehat{\psi}^3|\widehat{p_\shortparallel}|^5 \widehat{\psi} + \widehat{\psi}|\widehat{p_\shortparallel}|^5 \widehat{\psi}^3 \big) - \frac{1}{8}\big(\widehat{\psi}^2|\widehat{p_\shortparallel}|^5 \widehat{\psi}^2 \big) \\ \nonumber
    \widehat{K_T}^{(4)}&= \frac{1}{6} \big( \widehat{\psi}^3 (\partial_\alpha \widehat{\psi}) \hat{\partial_\alpha} |\hat{p_\shortparallel}|^3- (\partial_\alpha \widehat{\psi}) \hat{\partial_\alpha} |\hat{p_\shortparallel}|^3|\widehat{\psi}^3 \big) \\ \nonumber
    &+ \frac{1}{2} \big( \widehat{\psi} (\partial_\alpha \widehat{\psi}) \hat{\partial_\alpha} |\hat{p_\shortparallel}|^3 \widehat{\psi} ^2- \widehat{\psi} ^2(\partial_\alpha \widehat{\psi}) \hat{\partial_\alpha} |\hat{p_\shortparallel}|^3\widehat{\psi} \big)\\ \nonumber
    \widehat{K_U}^{(4)}&= \frac{1}{4}\big(\widehat{\psi}^2(\partial_\alpha \widehat{\psi}) \hat p_\alpha |\hat{p}_\shortparallel| \hat p_\beta (\partial_\beta \widehat{\psi})\big) + \frac{1}{4}\big((\partial_\alpha \widehat{\psi}) \hat p_\alpha |\hat{p}_\shortparallel| \hat p_\beta (\partial_\beta \widehat{\psi}) \widehat{\psi}^2\big) \\ \nonumber
    &-\frac{1}{2}\big(\widehat{\psi}(\partial_\alpha \widehat{\psi}) \hat p_\alpha |\hat{p}_\shortparallel| \hat p_\beta (\partial_\beta \widehat{\psi}) \widehat{\psi}\big) \,,
\end{align}
where we have used the symmetry $x_\shortparallel \to x'_\shortparallel$ 
(when present) beforehand to group operators that give the same 
trace when multiplied by $[\widehat{K}^{(0)}]^{-1}$. This leads to the 
kernels $\Gamma^{(4,1)}_S$,  $\Gamma^{(4,1)}_T$, and  $\Gamma^{(4,1)}_U$ 
respectively.

Following the steps of the second order calculation, we introduce the 
expression for the Dirichlet equivalent to $\Gamma^{(4,1)}$, namely, 
$\Gamma^{(4,1)}_D$, as

\begin{align}
    \Gamma^{(4,1)}_D= -\frac{1}{12} {\rm Tr} \big[|\widehat{p}_\shortparallel| \big(\widehat{\psi}^3|\widehat{p}_\shortparallel|^3 \widehat{\psi} + \widehat{\psi}|\widehat{p}_\shortparallel|^3 \widehat{\psi}^3 \big) \big]+\frac{1}{8} {\rm Tr} \big[|\widehat{p}_\shortparallel|\widehat{\psi}^2||\widehat{p}_\shortparallel|^3 \widehat{\psi}^2\big]\,,
\end{align}
having discarded  terms that do not contribute to the imaginary part. This 
allows us to write
\begin{align}
    \Gamma^{(4,1)}_S-\Gamma^{(4,1)}_D&= -\frac{1}{12} {\rm Tr} \big[|\widehat{p}_\shortparallel|^{-1} \widehat{\psi}^3|\widehat{p}_\shortparallel|^3 [\widehat{p}_\shortparallel^2\,,\,\widehat{\psi}] + |\widehat{p}_\shortparallel|^{-1}[\widehat{\psi}\, , \,\widehat{p}_\shortparallel^{2}]|\widehat{p}_\shortparallel|^3 \widehat{\psi}^3 \big] \\ \nonumber
    &+\frac{1}{8} {\rm Tr} \big[  |\widehat{p}_\shortparallel|^{-1}[\widehat{\psi}^2\, , \,\widehat{p}_\shortparallel^{2}]|\widehat{p}_\shortparallel|^3 \widehat{\psi}^2\big]\,.
\end{align}

Additionally, we have 

\begin{align}
    \Gamma^{(4,1)}_T&=\frac{1}{6} {\rm Tr}\big[ |\widehat{p}_\shortparallel|^{-1} (\partial_\alpha \widehat{\psi}) \hat{\partial_\alpha} |\widehat{p}_\shortparallel|^3 \widehat{\psi}^3\big] - \frac{1}{2} {\rm Tr}\big[ |\widehat{p}_\shortparallel|^{-1}\widehat{\psi} (\partial_\alpha \widehat{\psi}) \hat{\partial_\alpha} |\widehat{p}_\shortparallel|^3 \widehat{\psi}^2\big] \\ \nonumber 
    &+\frac{1}{2}{\rm Tr}\big[ |\widehat{p}_\shortparallel|^{-1}\widehat{\psi}^2 (\partial_\alpha \widehat{\psi}) \hat{\partial_\alpha} |\widehat{p}_\shortparallel|^3 \widehat{\psi}\big] \, ,
\end{align}
and

\begin{align}
    \Gamma_U^{(4,1)}&=-\frac{1}{2}{\rm Tr}\big[ |\widehat{p}_\shortparallel|^{-1} \widehat{\psi}^2 (\partial_\alpha \widehat{\psi}) \hat p_\alpha |\hat{p}_\shortparallel| \hat p_\beta  (\partial_\beta \widehat{\psi})\big]  \\ \nonumber
    &+\frac{1}{2}{\rm Tr}\big[|\widehat{p}_\shortparallel|^{-1} \widehat{\psi}(\partial_\alpha \widehat{\psi}) \hat p_\alpha |\hat{p}_\shortparallel| \hat p_\beta  (\partial_\beta \widehat{\psi})\widehat{\psi}\big] \,.
\end{align}

Using $[\widehat \psi^2,\widehat{p}^2]=\{\widehat{\psi}, [\widehat \psi, \widehat p^2] \}$, and $[\widehat \psi, \widehat p^2]=(\partial_\alpha \widehat \psi) \hat{\partial_\alpha} + \hat{\partial_\alpha} (\partial_\alpha \widehat{\psi})$, we can write

\begin{align} 
    \delta\,\Gamma^{(4,1)}=& \Gamma^{(4,1)}(\psi)- \Gamma^{(4,1)}_D(\psi)\\ \nonumber
    =&-\frac{1}{6} {\rm Tr}\big[ |\widehat{p}_\shortparallel|^{-1} \hat{\partial_\alpha} (\partial_\alpha \widehat \psi) |\widehat{p}_\shortparallel|^{3} \widehat{\psi}^3\big] + \frac{1}{8} {\rm Tr} \big[  |\widehat{p}_\shortparallel|^{-1}\{\widehat{\psi}, [\widehat \psi, \widehat p^2] \}|\widehat{p}_\shortparallel|^3  \widehat{\psi}^2\big] \\ \nonumber 
    &-\frac{1}{2} {\rm Tr} \big[  |\widehat{p}_\shortparallel|^{-1} \widehat \psi (\partial_\alpha \widehat \psi) |\widehat{p}_\shortparallel|^{3}\hat{\partial_\alpha}\widehat \psi^2\big] +\frac{1}{2} {\rm Tr} \big[  |\widehat{p}_\shortparallel|^{-1} \widehat \psi^2 (\partial_\alpha \widehat \psi) |\widehat{p}_\shortparallel|^{3}\hat{\partial_\alpha}\widehat \psi\big] \\ \nonumber
    &-\frac{1}{2}{\rm Tr} \big[  |\widehat{p}_\shortparallel|^{-1}\widehat{\psi}^2(\partial_\alpha \widehat{\psi}) \hat p_\alpha |\hat{p}_\shortparallel| \hat p_\beta (\partial_\beta \widehat{\psi})\big] +\frac{1}{2}{\rm Tr} \big[  |\widehat{p}_\shortparallel|^{-1}\widehat{\psi}(\partial_\alpha \widehat{\psi}) \hat p_\alpha |\hat{p}_\shortparallel| \hat p_\beta (\partial_\beta \widehat{\psi})\widehat{\psi}\big] \,.
\end{align}

After some algebra, this can be written as

\begin{align}\label{def delta 4,1}
    \delta\,\Gamma^{(4,1)}&= {\rm Tr}\big[ |\widehat{p}_\shortparallel|^{-1} \widehat{\psi}^2  (\partial_\alpha \widehat \psi) |\widehat{p}_\shortparallel|^{3} (\delta_{\alpha \beta}-\frac{1}{2}\frac{\widehat p_\alpha \widehat p_\beta}{|\widehat{p}_\shortparallel|^{2}}) (\partial_\beta \widehat \psi)\big] \\ \nonumber
    &-{\rm Tr}\big[ |\widehat{p}_\shortparallel|^{-1} \widehat{\psi}  (\partial_\alpha \widehat \psi) |\widehat{p}_\shortparallel|^{3} (\delta_{\alpha \beta}-\frac{1}{2}\frac{\widehat p_\alpha \widehat p_\beta}{|\widehat{p}_\shortparallel|^{2}}) (\partial_\beta \widehat \psi) \widehat{\psi}  \big] \\ \nonumber
    &+\frac{1}{6} {\rm Tr}\big[ |\widehat{p}_\shortparallel|^{-1} \widehat{\psi}^3 |\widehat{p}_\shortparallel|^{3} \hat{\partial_\alpha} (\partial_\alpha \widehat \psi)\big] - \frac{1}{2} {\rm Tr}\big[ |\widehat{p}_\shortparallel|^{-1}\hat{\partial_\alpha}  \widehat{\psi} |\widehat{p}_\shortparallel|^{3} (\partial_\alpha \widehat \psi)\widehat{\psi} ^2\big] \\ \nonumber
    &-\frac{1}{4}{\rm Tr}\big[ |\widehat{p}_\shortparallel|^{-1}\,\,(\hat{\partial_\alpha}  \widehat{\psi} (\partial_\alpha \widehat \psi)|\widehat{p}_\shortparallel|^{3} \widehat{\psi} ^2 + \widehat{\psi} ^2|\widehat{p}_\shortparallel|^{3}\hat{\partial_\alpha}  \widehat{\psi} (\partial_\alpha \widehat \psi) \,)\big] \,.
\end{align}

For simplicity, and in order to be able to write more explicit 
expressions, let us analyze the representative case of a wavelike surface:
\begin{align}
      \psi(x_\shortparallel)=2A\, {\rm cos}(\omega_0 \, x_0 - 
      \bm{\omega}_\shortparallel \cdot \bm{x}_\shortparallel),
\end{align}
which simplifies the outer-momentum structure of the system by imposing 

\begin{align} \label{Dynamic surface Fourier transform powers}
\tilde{\psi}(k_\shortparallel)&=A\,(2\pi)^d\left[\delta^d(k_\shortparallel-\omega_\shortparallel)+\delta^d(k_\shortparallel+\omega_\shortparallel)\right] \\ \nonumber
\tilde{\psi^2}(k_\shortparallel)&=A^2(2\pi)^d\left[\delta^d(k_\shortparallel-2\omega_\shortparallel)+\delta^d(k_\shortparallel+2\omega_\shortparallel)+2\delta^d(k_\shortparallel)\right] \\ \nonumber
\tilde{\psi^3}(k_\shortparallel)&= A^3(2\pi)^d\left[\delta^d(k_\shortparallel-3\omega_\shortparallel)+\delta^d(k_\shortparallel+3\omega_\shortparallel)+3\delta^d(k_\shortparallel-\omega_\shortparallel) + 3\delta^d(k_\shortparallel+\omega_\shortparallel)\right].
\end{align}

Taking the traces in \eqref{def delta 4,1}, this is reduced to

\begin{align}
    \delta\, \Gamma^{(4,1)}(\omega_\shortparallel,A)= A^4 (2\pi)^d \delta^d(0) \int_{\not \ell} \omega_\alpha \omega_\beta (\delta_{\alpha \beta }\;\ell_\shortparallel^2-\ell_\alpha \ell_\beta) \big( \frac{|\ell_\shortparallel|}{|\ell_\shortparallel-\omega_\shortparallel|} - \frac{|\ell_\shortparallel|}{|\ell_\shortparallel-2\omega_\shortparallel|} \big) \,,
\end{align}
which vanishes for $d=1$. Specifically

\begin{align}
    \frac{\delta\, \Gamma^{(4,1)}(\omega_\shortparallel,A)}{A^4 (2\pi)^d \delta^d(0)}=  \; (4^{\frac{d+2}{2}}-1) (d-1)
    \frac{\Gamma\!\left(-\tfrac{d}{2}-1\right)\,\Gamma\!\left(\frac{d+1}{2}\right)\,\Gamma\left(\frac{d+3}{2}\right)}
{(4\pi)^{\frac{d+2}{2}}\,\Gamma(d+2)}\, (\omega_\shortparallel^2)^{\frac{d+4}{2}}\,.
\end{align}

We can perform a Wick rotation $\omega_\shortparallel^2 \to -\omega^2$ to define $\Gamma^{(4,1)}_M(\omega,A)$ as the Minkowski counterpart of $\Gamma^{(4,1)}(\omega_\shortparallel,A)$, and write

    \begin{align}
    \frac{\text{Im}\big(\Gamma_M^{(4,1)}(\omega,A)\big)}{TV}=\sigma_N \, A^4  \, (\omega^2)^{q+\frac{5}{2}} \Theta(|\omega_0|-|\vec{\omega|})\,
\end{align}
and 

  \begin{align}
    \frac{\text{Im}\big(\Gamma_M^{(4,1)}(\omega_\shortparallel,A)\big)}{TV}=\kappa_N \, A^4  \, (\omega^2)^{q+2} \, \Theta(|\omega_0|-|\vec{\omega|}) , 
\end{align}
for odd $(d=2q+1)$ and even $(d=2q)$ dimensions respectively, where $TV=(2\pi)^d \delta^d(0)$. Similarly to the second order, for odd $d$ the imaginary part comes from the non-analytical momentum expression $(\omega^2)^{1/2}$, while for even $d$ it comes from a logarithmic term using dimensional regularization. The successions $\sigma_N$ and $\kappa_N$ are compared to their Dirichlet counterparts, $\sigma_D$ and $\kappa_D$, respectively, in figures \ref{fig:odd_ratio_order_4,1} and \ref{fig:even_ratio_order_4,1}.

\begin{figure}[h!] 
     \centering
     \includegraphics[width=0.7\textwidth]{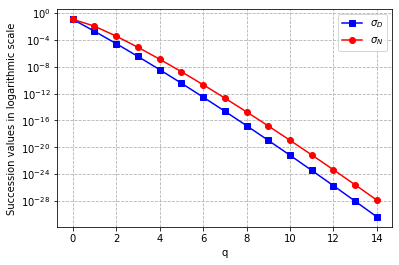}
     \caption{Successions $\sigma_D$ and $\sigma_N$ in logarithmic scale.}
     \label{fig:odd_ratio_order_4,1}
\end{figure}

\begin{figure}[h!] 
     \centering
     \includegraphics[width=0.7\textwidth]{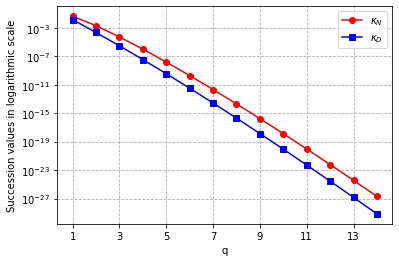}
     \caption{Successions $\kappa_D$ and $\kappa_N$ in logarithmic scale.}
     \label{fig:even_ratio_order_4,1}
\end{figure}

\subsubsection{$\Gamma^{(4,2)}$}

For this contribution, we have

\begin{align}
    \Gamma^{(4,2)}(\psi)&=-\frac{1}{4}{\rm Tr}\big[[K^{(0)}]^{-1}K^{(2)} [K^{(0)}]^{-1}K^{(2)}\big]\\ \nonumber
    &=-{\rm Tr}\big[|\widehat{p_\shortparallel}|^{-1} \widehat{K}_S^{(2)}|\widehat{p_\shortparallel}|^{-1} \widehat{K}_S^{(2)}\big] -{\rm Tr}\big[|\widehat{p_\shortparallel}|^{-1} \widehat{K}_T^{(2)}|\widehat{p_\shortparallel}|^{-1} \widehat{K}_T^{(2)}\big]\\ \nonumber
    &-{\rm Tr}\big[|\widehat{p_\shortparallel}|^{-1} \widehat{K}_U^{(2)}|\widehat{p_\shortparallel}|^{-1} \widehat{K}_U^{(2)}\big] \\ \nonumber
    &-2\,{\rm Tr}\big[|\widehat{p_\shortparallel}|^{-1} \widehat{K}_S^{(2)}|\widehat{p_\shortparallel}|^{-1} \widehat{K}_T^{(2)}\big]-2\, {\rm Tr}\big[|\widehat{p_\shortparallel}|^{-1} \widehat{K}_S^{(2)}|\widehat{p_\shortparallel}|^{-1} \widehat{K}_U^{(2)}\big]\\ \nonumber
    &-2\,{\rm Tr}\big[|\widehat{p_\shortparallel}|^{-1} \widehat{K}_T^{(2)}|\widehat{p_\shortparallel}|^{-1} \widehat{K}_U^{(2)}\big]\,,
\end{align}
which we'll write as

\begin{align}
    \Gamma^{(4,2)}(\psi)=\Gamma^{(4,2)}_{SS}+\Gamma^{(4,2)}_{TT}+\Gamma^{(4,2)}_{UU}+2\Gamma^{(4,2)}_{ST}+2\Gamma^{(4,2)}_{SU}+2\Gamma^{(4,2)}_{TU}
\end{align}
in an obvious notation.

We start with

\begin{align}
    \Gamma^{(4,2)}_{SS}&=-{\rm Tr}\big[[K^{(0)}]^{-1}K_S^{(2)} [K^{(0)}]^{-1}K_S^{(2)}\big] \\ \nonumber&=-\frac{1}{8} {\rm Tr}\big[|\widehat{p_\shortparallel}|^{-1} \widehat \psi^2 |\widehat{p_\shortparallel}|^{5} \widehat \psi^2\big] + \frac{1}{2} {\rm Tr}\big[|\widehat{p_\shortparallel}|^{-1} \widehat \psi |\widehat{p_\shortparallel}|^{3} \widehat \psi  |\widehat{p_\shortparallel}|^{2} \widehat\psi^2\big] \\ \nonumber
    &-\frac{1}{4}{\rm Tr}\big[|\widehat{p_\shortparallel}|^{-1} \widehat \psi |\widehat{p_\shortparallel}|^{3} \widehat \psi  \, |\widehat{p_\shortparallel}|^{-1} \widehat \psi |\widehat{p_\shortparallel}|^{3} \widehat \psi \big]\, ,
\end{align}
and, following the previous procedure, we introduce the Dirichlet counterpart 

\begin{align}
    \Gamma^{(4,2)}_D&=-\frac{1}{8} {\rm Tr}\big[|\widehat{p_\shortparallel}| \widehat \psi^2 |\widehat{p_\shortparallel}|^{3} \widehat \psi^2\big] + \frac{1}{2} {\rm Tr}\big[|\widehat{p_\shortparallel}| \widehat \psi |\widehat{p_\shortparallel}| \widehat \psi  |\widehat{p_\shortparallel}|^{2} \widehat\psi^2\big] \\ \nonumber
    &-\frac{1}{4}{\rm Tr}\big[|\widehat{p_\shortparallel}|\widehat \psi |\widehat{p_\shortparallel}| \widehat \psi  \, |\widehat{p_\shortparallel}| \widehat \psi |\widehat{p_\shortparallel}| \widehat \psi \big]\, .
\end{align}

This allows us to write

\begin{align} \label{def SS-D 4,2}
    \Gamma^{(4,2)}_{SS}-\Gamma^{(4,2)}_D&= -\frac{1}{8} {\rm Tr}\big[|\widehat{p_\shortparallel}|^{-1} [\widehat \psi^2 , \widehat{p_\shortparallel}^{2}] |\widehat{p_\shortparallel}|^3\widehat \psi^2\big] + \frac{1}{2} {\rm Tr}\big[|\widehat{p_\shortparallel}|^{-1} [\widehat \psi ,  \widehat{p_\shortparallel}^2] |\widehat{p_\shortparallel}| \widehat \psi  |\widehat{p_\shortparallel}|^{2} \widehat\psi^2\big] \\ \nonumber
    &-\frac{1}{2}{\rm Tr}\big[|\widehat{p_\shortparallel}|^{-1} [\widehat \psi ,  \widehat{p_\shortparallel}^2] |\widehat{p_\shortparallel}| \widehat \psi |\widehat{p_\shortparallel}| \widehat \psi |\widehat{p_\shortparallel}| \widehat \psi\big]\\ \nonumber
    &-\frac{1}{4}{\rm Tr}\big[|\widehat{p_\shortparallel}|^{-1} [\widehat \psi ,  \widehat{p_\shortparallel}^2] |\widehat{p_\shortparallel}| \widehat \psi |\widehat{p_\shortparallel}|^{-1} [\widehat \psi ,  \widehat{p_\shortparallel}^2] |\widehat{p_\shortparallel}| \widehat \psi \big]\nonumber \,.
\end{align}
Furthermore, we have

\begin{align}
     \Gamma^{(4,2)}_{TT}&= {\rm Tr} \big[ |\widehat{p_\shortparallel}|^{-1} \widehat \psi  (\partial_\alpha \widehat \psi) \hat\partial_\alpha (\partial_\beta \widehat \psi) |\widehat{p_\shortparallel}| \hat\partial_\beta \widehat \psi\big]-{\rm Tr} \big[ |\widehat{p_\shortparallel}|^{-1} (\partial_\beta \widehat \psi) |\widehat{p_\shortparallel}| \hat\partial_\beta \widehat \psi \hat\partial_\alpha (\partial_\alpha \widehat \psi)    \widehat \psi\big] \\ \nonumber 
     &+\frac{1}{2}{\rm Tr} \big[ |\widehat{p_\shortparallel}|^{-1} \widehat \psi  (\partial_\alpha \widehat \psi) |\widehat{p_\shortparallel}|  \hat\partial_\alpha \hat\partial_\beta  (\partial_\beta \widehat \psi) \widehat \psi\big] \\ \nonumber
     &+\frac{1}{2}{\rm Tr} \big[ |\widehat{p_\shortparallel}|^{-1} \widehat \psi |\widehat{p_\shortparallel}|  \hat\partial_\beta  (\partial_\beta \widehat \psi) |\widehat{p_\shortparallel}|^{-1} (\partial_\alpha \widehat \psi) |\widehat{p_\shortparallel}|  \hat\partial_\alpha \widehat \psi\big] \\ \nonumber
     &-\frac{1}{2}{\rm Tr} \big[ |\widehat{p_\shortparallel}|^{-1} \widehat \psi |\widehat{p_\shortparallel}|  \hat\partial_\beta  (\partial_\beta \widehat \psi) |\widehat{p_\shortparallel}|^{-1} \widehat \psi |\widehat{p_\shortparallel}|  \hat\partial_\alpha (\partial_\alpha \widehat \psi) \big] \, ,
\end{align}
and 
\begin{align}
    \Gamma^{(4,2)}_{UU}&= -\frac{1}{4} {\rm Tr} \big[ |\widehat{p_\shortparallel}|^{-1}(\partial_\alpha \widehat \psi) \frac{\widehat p_\alpha \widehat p_\beta}{|\widehat{p_\shortparallel}|}(\partial_\beta \widehat \psi) |\widehat{p_\shortparallel}|^{-1}(\partial_\gamma \widehat \psi) \frac{\widehat p_\gamma \widehat p_\sigma}{|\widehat{p_\shortparallel}|}(\partial_\sigma \widehat \psi)\big]\,.
\end{align}

For the non-diagonal terms, we have

\begin{align}
     2\Gamma^{(4,2)}_{ST}&= \frac{1}{2}{\rm Tr} \big[ |\widehat{p_\shortparallel}|^{-1} \widehat \psi |\widehat{p_\shortparallel}|\hat\partial_\alpha (\partial_\alpha \widehat \psi) |\widehat{p_\shortparallel}|^2 \widehat \psi^2 \big]+\frac{1}{2}{\rm Tr} \big[ |\widehat{p_\shortparallel}|^{-1} \widehat \psi^2 |\widehat{p_\shortparallel}|^2 \widehat \psi |\widehat{p_\shortparallel}|\hat\partial_\alpha (\partial_\alpha \widehat \psi)  \big] \\ \nonumber 
     &-\frac{1}{2}{\rm Tr} \big[ |\widehat{p_\shortparallel}|^{-1} \widehat \psi^2 |\widehat{p_\shortparallel}|^3 \hat\partial_\alpha (\partial_\alpha \widehat \psi) \widehat \psi\big] + {\rm Tr} \big[ |\widehat{p_\shortparallel}|^{-1} \widehat \psi |\widehat{p_\shortparallel}|^3 \widehat \psi \hat\partial_\alpha (\partial_\alpha \widehat \psi) \widehat \psi\big]\\ \nonumber
     &-{\rm Tr} \big[ |\widehat{p_\shortparallel}|^{-1} \widehat \psi |\widehat{p_\shortparallel}|^3 \widehat \psi |\widehat{p_\shortparallel}|^{-1} \widehat \psi |\widehat{p_\shortparallel}| \hat\partial_\alpha (\partial_\alpha \widehat \psi) \big] \, ,
\end{align}
\begin{align}
     2\Gamma^{(4,2)}_{SU}&= \frac{1}{2}{\rm Tr} \big[ |\widehat{p_\shortparallel}|^{-1} \widehat \psi^2|\widehat{p_\shortparallel}|^{2} (\partial_\alpha \widehat \psi) \frac{\widehat p_\alpha \widehat p_\beta}{|\widehat{p_\shortparallel}|}(\partial_\beta \widehat \psi)\big] \\ \nonumber
     &-\frac{1}{2} {\rm Tr} \big[ |\widehat{p_\shortparallel}|^{-1} \widehat \psi|\widehat{p_\shortparallel}|^{3}\widehat \psi |\widehat{p_\shortparallel}|^{-1}(\partial_\alpha \widehat \psi) \frac{\widehat p_\alpha \widehat p_\beta}{|\widehat{p_\shortparallel}|}(\partial_\beta \widehat \psi) \big] 
\end{align}
and
\begin{align} \label{def TU 4,2}
     2\Gamma^{(4,2)}_{TU}&= -{\rm Tr} \big[ |\widehat{p_\shortparallel}|^{-1} \widehat \psi (\partial_\alpha \widehat \psi) \hat\partial_\alpha (\partial_\gamma \widehat \psi) \frac{\widehat p_\gamma \widehat p_\sigma}{|\widehat{p_\shortparallel}|}(\partial_\sigma \widehat \psi)\big] \\ \nonumber
     &+{\rm Tr} \big[ |\widehat{p_\shortparallel}|^{-1}  (\partial_\alpha \widehat \psi)|\widehat{p_\shortparallel}| \hat\partial_\alpha \widehat \psi |\widehat{p_\shortparallel}|^{-1}(\partial_\gamma \widehat \psi) \frac{\widehat p_\gamma \widehat p_\sigma}{|\widehat{p_\shortparallel}|}(\partial_\sigma \widehat \psi)\big] \,.
\end{align}

Taking the traces for the case where $\psi$ is defined by \eqref{Dynamic surface Fourier transform powers}, we identify three different structures that will be presented below. In general, we have-two propagator integrals of the forms

\begin{align}
    I_1(\omega_\shortparallel)= \int_{\not \ell} \frac{F_1(\ell_\shortparallel, \omega_\shortparallel)}{|\ell_\shortparallel| |\ell_\shortparallel + \omega_\shortparallel|}
\end{align}
and 
\begin{align}
    I_2(\omega_\shortparallel)= \int_{\not \ell} \frac{F_2(\ell_\shortparallel, \omega_\shortparallel)}{|\ell_\shortparallel| |\ell_\shortparallel +2 \omega_\shortparallel|} \, ,
\end{align}
where $F_1$ and $F_2$ are functions of $\ell_\shortparallel^2$, $\omega_\shortparallel^2$, and $\omega_\shortparallel \cdot \ell_\shortparallel$, of degree $6$ in momentum powers. 
The third structure that appears is a three-propagator integral of the form

\begin{align}
    I_3(\omega_\shortparallel)= \int_{\not \ell} \frac{F_3(\ell_\shortparallel, \omega_\shortparallel)}{|\ell_\shortparallel|^2|\ell_\shortparallel+\omega_\shortparallel| |\ell_\shortparallel - \omega_\shortparallel|} \, ,
\end{align}
where $F_3$ is also a function of of $\ell_\shortparallel^2$, $\omega_\shortparallel^2$, and $\omega_\shortparallel \cdot \ell_\shortparallel$, but it has a momentum-power degree of $8$. All integrals that appear in this calculation can be put into the form of either $I_1$, $I_2$, or $I_3$ by multiplying and dividing by $|\ell_\shortparallel|$, $|\ell_\shortparallel\pm\omega_\shortparallel|$, and $|\ell_\shortparallel\pm 2\omega_\shortparallel|$, absorbing all momentum-squared factors into $F_1$, $F_2$, or $F_3$.

Following this procedure for the traces from \eqref{def SS-D 4,2} to \eqref{def TU 4,2}, we can write

\begin{align}
    \delta\, \Gamma^{(4,2)} (\omega_\shortparallel,A) &= \Gamma^{(4,2)}(\omega_\shortparallel,A)-\Gamma_D^{(4,2)}(\omega_\shortparallel,A)\\ \nonumber 
    &=A^4 (2\pi)^d \delta^d(0) \big(I_1(\omega_\shortparallel) +I_2(\omega_\shortparallel)+I_3(\omega_\shortparallel)\big) \, ,
\end{align}
with

\begin{align} \label{def: I_1}
    I_1(\omega_\shortparallel)= \int_{\not \ell} \frac{5\ell_\shortparallel^2 (\ell_\shortparallel \cdot \omega_\shortparallel)^2 - 5 \ell_\shortparallel^4 \omega_\shortparallel^2+\frac{9}{4} (\ell_\shortparallel \cdot \omega_\shortparallel)^2 \omega_\shortparallel^2 - \frac{9}{4} \ell_\shortparallel ^2 \omega_\shortparallel^4}{|\ell_\shortparallel+\frac{\omega_\shortparallel}{2}| |\ell_\shortparallel - \frac{\omega_\shortparallel}{2}|} \, ,
\end{align}

\begin{align} \label{def: I_2}
    I_2(\omega_\shortparallel)= \int_{\not \ell} \frac{\ell_\shortparallel^4 \omega_\shortparallel^2-2\ell_\shortparallel^2 (\ell_\shortparallel \cdot \omega_\shortparallel)^2 - \frac{1}{2} \ell_\shortparallel ^2 \omega_\shortparallel^4+ 3 (\ell_\shortparallel \cdot \omega_\shortparallel)^2 \omega_\shortparallel^2 -\omega_\shortparallel^6}{|\ell_\shortparallel+\omega_\shortparallel| |\ell_\shortparallel-\omega_\shortparallel|} \, ,
\end{align}
and
\begin{align} \label{def: I_3}
    I_3(\omega_\shortparallel)= \int_{\not \ell} \frac{-\ell_\shortparallel^4 (\ell_\shortparallel \cdot \omega_\shortparallel)^2 - (\ell_\shortparallel \cdot \omega_\shortparallel)^4 +2\ell_\shortparallel^6 \omega_\shortparallel^2-4\ell_\shortparallel^2 (\ell_\shortparallel \cdot \omega_\shortparallel)^2\omega_\shortparallel^2+\frac{5}{2}\ell_\shortparallel^4 \omega_\shortparallel^4  + \ell_\shortparallel^2\omega_\shortparallel^6}{|\ell_\shortparallel|^2|\ell_\shortparallel+\omega_\shortparallel| |\ell_\shortparallel-\omega_\shortparallel|} \, ,
\end{align}
where the denominators have been put in a symmetric form  so that we can 
discard terms of odd power in $\ell_\shortparallel$.

Notice that all the 
terms of $I_3(\omega_\shortparallel)$, except the one with 
$(\ell_\shortparallel \cdot \omega_\shortparallel)^4$, could be absorbed 
into $I_2(\omega_\shortparallel)$ by canceling the 
$|\ell_\shortparallel|^2$ propagator in the denominator. However, because 
of the presence of that term, the quantity $\delta\, \Gamma^{(4,2)}$ 
depends on a three-point one-loop function.  They have been studied for 
specific mass and momentum configurations 
in~\cite{Suzuki2002,PhanTran2019}. Naturally, they are considerably more 
involved than two-propagator integrals, which makes obtaining a numerical 
result for general $d$ challenging. However, for $d=1$, we have 
$(\ell_\shortparallel \cdot 
\omega_\shortparallel)^{2n} = (|\ell_\shortparallel|^2 
|\omega_\shortparallel|^2)^{n}$ for integer $n$, so \eqref{def: I_3} can 
be absorbed completely in the structure of \eqref{def: I_2}. With this 
condition, we find $I_1(\omega_\shortparallel)=0$ and 
$I_2(\omega_\shortparallel)+I_3(\omega_\shortparallel)=0$, which means 
$\delta \, \Gamma^{(4,2)}(\omega_\shortparallel,A)=0$ for $d=1$. Since we 
also had $\delta \, \Gamma^{(4,1)}(\omega_\shortparallel,A)=0$ for $d=1$, 
this means that $d=1$ implies $\delta \, 
\Gamma^{(4)}(\omega_\shortparallel,A)=0$.

Thus, we have shown explicitly  that, up to the fourth order, the Neumann 
result agrees for $d=1$ with the Dirichlet one for wavelike surfaces, and 
found the explicit form of the difference for other dimensions in terms of general one-loop integrals.

It's worth mentioning that naively performing the loop integrals without 
first looking for cancellations in $d=1$ would lead to troublesome 
expressions involving $\Gamma(d-1)$ and $\Gamma(\frac{d-1}{2})$, for both 
the second and fourth orders. In $d=1$, those Gamma functions are 
divergent, and using dimensional regularization $d=1-\epsilon$ produces 
non-analytic counter-terms. However, we have verified for $\Gamma^{(2)}$ 
and $\Gamma^{(4,1)}$ that using an auxiliary mass $m$ for the scalar field 
and taking the limit $m \to 0$ after performing the integrals yields the 
same results as in this work, $d=1$ included.

This apparent divergence can alsa be seen to emerge if one uses the 
integration by parts (IBP) relations~\cite{ChetyrkinTkachov,Laporta}, in 
order to express loop integrals of the Neumann contributions as functions 
of known Dirichlet contributions. Specifically, if we define
\begin{equation}
   I(\lambda_1,\lambda_2)= \int_{p_\shortparallel} \frac{1}{(p_\shortparallel^2)^{\lambda_1} [(p_\shortparallel+k_\shortparallel)^2]^{\lambda_2}}\,,
\end{equation}
and consider $I(-\frac{1}{2},-\frac{1}{2})$ as our master integral, which 
is the Kernel for the Dirichlet case, we can use the IBP relations to 
solve for one of the Kernels of the Neumann case, namely 
$I(\frac{1}{2},-\frac{3}{2})$, as   $I(\frac{1}{2},-\frac{3}{2})= 
3\frac{d+1}{d-1}\;I(-\frac{1}{2},-\frac{1}{2})$, which brings problems for 
$d=1$ as mentioned. However, $I(\frac{1}{2},-\frac{3}{2})$ is not the only 
Kernel for the Neumann case, and the divergences  cancel when adding all 
the contributions.

\section{Conclusions}\label{sec:conc}
We have calculated the imaginary part of the effective action, for a 
Neumann surface in $d+1$ dimensions that can deform and move, in a 
time-dependent fashion. The approach relies upon a previous result for a 
Dirichlet surface, since we have managed to find explicit expressions for the difference between the imaginary parts for both 
kinds of surfaces.

We have assumed  small departures with respect to an average,  planar 
hypersurface, and performed an expansion up to the fourth order in the 
amplitude of the deformation.

Our results contribute, we believe, to the growing understanding of DCE's 
and motion-induced radiation phenomena. The explicit dimensional 
dependence provides benchmarks for future numerical and potentially 
experimental studies. The exponential suppression of dissipation with 
increasing dimension, suggests that experimental verification of these 
effects would be most feasible in low-dimensional systems, consistent with 
current experimental efforts focusing on quasi-one-dimensional cavity 
setups and two-dimensional systems.

Several extensions of this work are worth pursuing. First, the calculation 
could be extended to higher orders to examine whether the systematic 
difference between Neumann and Dirichlet conditions persists and to 
determine if there are more unexpected cancellations. 
Second, the  restriction to wavelike surfaces simplified the momentum 
structure considerably; a more general analysis of arbitrary surface 
deformations would provide deeper insight into the dependence on the 
surface's dynamical properties.  
Finally, investigating other boundary conditions, such as Robin or mixed 
conditions, would complete the picture of how quantum fields respond to 
different types of dynamical boundary conditions.

\section*{Acknowledgements}
This work was supported by ANPCyT, CONICET, and UNCuyo.

\end{document}